\documentclass[letterpaper,twocolumn,10pt]{article}

\usepackage{algpseudocode}
\usepackage{xspace}
\usepackage{multirow}
\usepackage{makecell}
\usepackage{graphicx,amssymb,amsmath,endnotes}
\usepackage{epstopdf}
\usepackage{ulem}
\usepackage[tight]{subfigure}
\usepackage{fancyhdr}
\usepackage{textcomp, libertine}
\usepackage{xcolor}
\usepackage{etoolbox}
\date{}
\definecolor{red}{rgb}{1,0,0}
\begin{document}
\title{A Case for Sampling Based Learning Techniques in Coflow Scheduling}
\author{Akshay Jajoo, Y. Charlie Hu, Xiaojun Lin}

\newcommand{\algcaption}[1]{%
  \par\noindent
  \textbf{Algorithm:}
  #1\par
}

\providetoggle{tons}
\settoggle{tons}{true}

\pagenumbering{gobble}
\newcommand{\oldstuff}[1]{}
\newcommand{\yes}{$\checkmark$}
\newcommand{\limited}{Limited}
\newcommand{\no}{$\times$}
\definecolor{gray}{rgb}{0.5,0.5,0.5}
\newcommand{\etc}{\emph{etc.}\xspace}
\newcommand{\etcc}{\emph{etc.}}
\newcommand{\ie}{\emph{i.e.,}\xspace}
\newcommand{\eg}{\emph{e.g.,}\xspace}
\newcommand{\etal}{\emph{et al.}\xspace}
\newcommand{\SmallCrunch}{\vspace{-0cm}}
\newcommand{\smallcrunch}{\vspace{-0cm}}

\renewcommand{\paragraph}[1]{\smallskip\noindent{\bf{#1}}}

\newcommand{\comment}[1]{{#1}}
\iftoggle{tons}
{
	\newcommand{\commenttons}[1]{{\textit{#1}}}
}
{
	\newcommand{\commenttons}[1]{{\color{red} \textit{}}}
}
\newcommand{\commentlinx}[1]{{\color{red} \textit{linx: #1}}}
\newcommand{\commentaj}[1]{{\color{red} \textit{aj: #1}}}
\newcommand{\questionaj}[1]{{\color{red} \textbf{questionAJ: #1}}}
\newcommand{\updated}[1]{{\color{red} {}}}
\newcommand{\editaj}[2]{{\sout{#1}\color{red} {#2}}}
\newcommand{\editlinx}[2]{{\sout{#1}\color{blue} {#2}}}
\newcommand{\removeaj}[2]{{\color{red}{\textbf{Remove following:}}}{\sout{#1}}	{\color{red} {\textbf{Because} #2}}}
\newcommand{\addaj}[1]{{\color{red} {ADD?: #1}}}
\newcommand{\addtonreview}[1]{{#1}}
\newcommand{\removetonreview}[2]{}
\newcommand{\edittonreview}[2]{{{#2}}}

\newcommand{\insitu}{{\em in situ}\xspace}

\newcommand{\name}{{\sc Philae}\xspace}
\newcommand{\namee}{{\sc Philae}}
\newcommand{\ychurm}[1]{{\hspace{-0.2cm}}}

\newcommand{\panic}[1]{\vspace{-#1 plus 1pt minus 1pt}}
\newcommand{\panictwo}[1]{\vspace{-#1 plus 2pt minus 2pt}}

\newcommand{\nsection}[1]{\panictwo{0pt}\section{#1}\panictwo{0pt}}
\newcommand{\nsubsection}[1]{\panictwo{0pt}\subsection{#1}\panictwo{0pt}}
\newcommand{\nsubsubsection}[1]{\panictwo{0pt}\subsubsection{#1}\panictwo{0pt}}

\newcommand{\todoaj}[1]{{\color{red} \textit{TODO(AJ): #1}}}
\newcommand{\flow}{flow\xspace}
\newcommand{\linx}[1]{{ #1}}

\newcommand{\nnone}{n_1}
\newcommand{\nntwo}{n_2}
\newcommand{\mone}{m_1}
\newcommand{\mtwo}{m_2}
\newcommand{\muone}{\mu_1}
\newcommand{\mutwo}{\mu_2}
\newcommand{\sigmaone}{\sigma_1}
\newcommand{\sigmatwo}{\sigma_2}
\newcommand{\Tc}{T^c}
\newcommand{\Tcprime}{\tilde{T}^c}
\newcommand{\coflowone}{C_1}
\newcommand{\coflowtwo}{C_2}
\newcommand{\deltaprime}{\tilde{\delta}}
\newcommand{\xone}{X_1}
\newcommand{\xtwo}{X_2}
\newcommand{\tone}{t_1}
\newcommand{\ttwo}{t_2}

\maketitle

\begin{abstract}
Coflow scheduling improves data-intensive application performance by improving
their networking performance. State-of-the-art online coflow schedulers in
essence approximate the classic Shortest-Job-First (SJF) scheduling by learning
the coflow \textit{size} online.  In particular, they use multiple priority
queues to simultaneously accomplish two goals: to sieve long coflows from short
coflows, and to schedule short coflows with high priorities.  Such a mechanism
pays high overhead in learning the coflow size: moving a large coflow across
the queues delays small and other large coflows, and moving similar-sized
coflows across the queues results in inadvertent round-robin scheduling.  We
propose \name, a new online coflow scheduler that exploits the spatial
dimension of coflows, \ie a coflow has many flows, to drastically reduce the
overhead of coflow size \textit{learning}.  \name pre-schedules sampled flows
of each coflow and uses their sizes to estimate the average flow size of the
coflow.  It then resorts to Shortest Coflow First, where the notion of shortest
is determined using the learned coflow sizes and coflow contention.  We show
that the sampling-based learning is robust to flow size skew and has the added
benefit of much improved scalability from reduced coordinator-local agent
interactions. Our evaluation using an Azure testbed, a publicly available
production cluster trace from Facebook shows that compared to the prior art
Aalo, \name reduces the coflow completion time (CCT) in average (P90) cases by
1.50$\times$ (8.00$\times$) on a 150-node testbed and 2.72$\times$
(9.78$\times$) on a 900-node testbed.  Evaluation using additional traces
further demonstrates \name's robustness to flow size skew.
\footnote{\comment{An earlier conference version of this work was presented at
USENIX ATC 2019~\cite{jajooPhilae}.}}


\end{abstract}

\vspace{-0.2in}
\section{Background and Problem Statement}
\label{sec:back}

We start with a brief review of the coflow abstraction and the need for
non-clairvoyant coflow scheduling and state the network model. \commenttons{We
then give an overview of existing online coflow schedulers and
formally state the problem.}


\paragraph{Coflow abstraction}
In data-parallel applications such as Hadoop~\cite{hadoop:web} and
Spark~\cite{spark:web}, the job completion time heavily depends on the
completion time of the communication stage~\cite{mantri:OSDI2010,
orchestra:sigcomm11}.  The coflow abstraction~\cite{coflow:hotnets12} was
proposed to speed up the communication stage to improve application
performance. A coflow is defined as a set of flows between several nodes that
accomplish a common task. For example, in map-reduce jobs, the set of all flows
from all map to all reduce tasks in a single job forms a typical coflow. The
coflow completion time (CCT) is defined as the time duration between when the
first flow arrives and the last flow completes. In such applications, improving
CCT is more important than improving individual flows' completion time (FCT)
for improving the application performance~\cite{aalo:sigcomm15,
varys:sigcomm14, baraat:sigcomm14, graviton:hotcloud16, jajooSaath}.

\paragraph{Non-clairvoyant coflows}
Data-parallel directed acyclic graphs (DAGs) typically have multiple stages
which are represented as multiple coflows with dependencies between them.
Recent systems (\eg~\cite{mapreduceonline:nsdi2010, apache:tez,
dryad:eurosys2007, dandelion:sosp2013}) employ optimizations that pipeline the
consecutive computation stages which removes the barrier at the end of each
coflow, making knowing flow sizes of each coflow beforehand difficult. 
A recent study~\cite{plausibleFlowSize} further shows various other reasons
why it is not very plausible to learn flow sizes from applications, for
example, learning flow sizes from applications requires changing either the
network stack or the applications.
Thus in this paper, we focus on {\em non-clairvoyant} coflow scheduling which do not
assume knowledge about coflow characteristics such as flow sizes upon coflow
arrival.

\paragraph{Non-blocking network fabric}
We assume the same non-blocking network fabric model in recent network
designs for coflows~\cite{varys:sigcomm14, aalo:sigcomm15,
  graviton:hotcloud16, jajooSaath, sincronia:sigcomm18}, where the
datacenter network fabric is abstracted as a single non-blocking
switch that interconnects all the servers, and each server
  (computing node) is abstracted as a network port that sends and
  receives flows. In such a model, the ports, \ie server uplinks and downlinks, are
the only source of contention as the network core is assumed to be
able to sustain all traffic injected into the network. We note that
the abstraction is to simplify our description and analysis, and is
not required or enforced in our evaluation.

\iftoggle{tons}{
{
\subsection{Prior-art on non-clairvoyant coflow scheduling}
\label{sec:back:aalo}


A classic approach to reduce average CCT is Shortest Coflow
First (SCF)~\cite{aalo:sigcomm15} (derived from classic SJF), where the coflow size is
loosely defined as the total bytes of the coflow, \ie sum of length of all its
flows. However, using SCF \textit{online} is not practical as it requires prior
knowledge about the coflow sizes. This is further complicated as coflows arrive
and exit dynamically and by other cluster dynamics such as failures and stragglers.

Aalo~\cite{aalo:sigcomm15} was proposed to schedule coflows online without any
prior knowledge.  The key idea in Aalo is to approximate SCF by learning Coflow
length
using {discrete priority queues}.
In particular, it starts a newly arrived coflow in the highest priority queue
and gradually moves it to the lower priority queues when the total data sent by
the coflow exceeds the per-queue thresholds.

The above idea {of learning the order of jobs in a priority queues was
originally applied to scheduling jobs on a single server}. To apply it to
scheduling coflows with many constituent flows over many network ports, \ie in
a distributed setting, Aalo uses a global coordinator to assign coflows to
logical priority queues, and uses the total bytes sent by all flows of a coflow
as its logical ``length'' in moving coflows across the queues.  The logical
priority queues are mapped to local priority queues at each port, and the
individual local ports act {\em independently} in scheduling flows in its local
priority queues, \eg by enumerating flows from the highest to lowest priority
queues and using FIFO to order the flows within each queue.

Generally speaking, using multiple priority queues in Aalo in this way has
three effects: (1) {\bf Coflow segregation:} It segregates long coflows (who
will move to low priority queues) from short coflows who will finish while in
high priority queues; (2) {\bf Finishing short coflows sooner:} Since high
priority queues receive more bandwidth allocation, short coflows will finish
sooner (than longer ones); (3) {\bf Starvation avoidance:} Using the FIFO
policy for intra-queue scheduling provides starvation avoidance, since at every
scheduling slot, each queue at each port receives a fixed bandwidth allocation
and FIFO ensures that every Coflow (its flow) in each queue is never moved
back.

Similar to Aalo~\cite{aalo:sigcomm15}, Graviton~\cite{graviton:hotcloud16} also
uses a logical priority queue structure.
Unlike Aalo, Graviton uses sophisticated policies to sort coflows
within a queue based on their width (total number of ports that a coflow is
present on).  
Saath~\cite{jajooSaath} is another priority queue based online coflow scheduler
that improves over Aalo with
three high-level design priniciples:
(1) it schedules flows of a coflow in an all-or-none fashion to prevent flows
of a coflow from going out-of-sync;
(2) it incorporates contention, \ie with how many other coflows a coflow is
sharing ports with, into the metric for sorting coflows within a queue;
(3) instead of using the total coflow size, it uses the length of the longest
flow of a coflow to determine transition across priority queues, which helps in
deciding the correct priority queue of a coflow faster.

}}{}
\vspace{-0.15in}
\removetonreview{\subsection{Motivation}

Using multiple priority queues in the above prior-art to implicitly
learn the relative sizes of coflows, however, negatively
affects the average CCT and the scalability of the coordinator in
several ways:

\edittonreview{
\textbf{(1) Intrinsic queue-transit overhead:} Every coflow that Aalo transits
through the queues before reaching its final queue worsens the average CCT because
during transitions, such a coflow effectively {\em blocks other shorter} coflows in the
earlier queues it went through, which would have been scheduled before this coflow starts in a
perfect SJF.


\textbf{(2) Overhead due to inadvertent round-robin:} Although Aalo
attempts to approximate SJF, it inadvertently ends up doing
{\em round-robin} for {coflows of similar sizes} as it moves them across
queues. Aalo assigns a \textit{fixed threshold of data transfer} 
for each queue. Assume there are ``$N$'' coflows in a queue
that do not finish in that queue. Aalo would schedule these coflows (following
FIFO) one by one and demote each of them to a lower priority queue when the coflow
reaches the queue threshold. 
Effectively, these 
coflows experience the round-robin scheduling which is known to have
the worst average {CCT}~\cite{OSConcepts}, when jobs are of similar sizes.


\textbf{(3) Limited scalability from frequent updates from local
ports:} To support the ``try-and-error'' style learning, the coordinator
requires frequent updates from all local ports of the bytes sent for
each coflow in order to move coflows across multiple queues
timely. This results in high load on the central coordinator from
receiving frequent updates and calculating and sending new rate
allocations, which limits the scalability of the overall approach.
}{
We are listing major three of them here which have been discussed in details in \S\ref{sec:intro:moti}
{(1) Intrinsic queue-transit overhead}
{(2) Overhead due to inadvertent round-robin}
{(3) Limited scalability from frequent updates from local
ports}
}
}

\subsection{Problem statement}

Our goal is to {\em develop an efficient non-clairvoyant coflow scheduler
that optimizes the communication performance, in particular the
average CCT, of data-intensive applications without prior knowledge,
while guaranteeing starvation freedom and work conservation and being
resilient to the network dynamics}. The problem of non-clairvoyant
coflow scheduling is NP-hard because coflow scheduling even assuming
all coflows arrive at time 0 and their size are known in advance is
already NP-hard~\cite{varys:sigcomm14}. Thus practical non-clairvoyant
coflow schedulers are approximation algorithms. Our approach is to
dynamically prioritize coflows by efficiently learning their flow
sizes online.

\section{Key Idea}
\label{sec:key}

Our new non-clairvoyant coflow scheduler design, \name, 
is based on a key observation about coflows that a coflow has a {\em  spatial
dimension}, \ie it typically consists of many flows.
We thus propose to explicitly learn coflow sizes online by using \textit{sampling}, a
highly effective technique used in large-scale surveys~\cite{samplingOfPopulations}. 
In particular, \name preschedules sampled flows, called {\em pilot flows},
of each coflow and uses their measured sizes to estimate the coflow
size. It then resorts to SJF or variations using the estimated coflow
sizes.

Developing a complete non-clairvoyant coflow scheduler based on the simple sampling idea raises
three questions:

{\em
(1) Why is sampling more efficient than the priority-queue-based coflow size learning? Would scheduling the remaining flows after sampled pilot flows are completed
adversely affect the coflow completion time?

(2) Will sampling be effective in the presence of skew of flow sizes?

(3) How to design the complete scheduler architecture?
}
We answer the first two questions below, and present the complete architecture design in \S\ref{sec:design}.

\vspace{-0.1in}
\subsection{\addtonreview{Why is sampling-based learning more efficient 
than priority-queue-based learning?}}
\label{subsec:why}

Scheduling pilot flows first before the rest of the flows can potentially incur
two sources of overhead.  First, scheduling pilot flows of a newly arriving
coflow consumes port bandwidth which can delay other coflows (with already
estimated sizes). However, compared to the multi-queue based approach, the
overhead is much smaller for two reasons: (1) \name schedules only a small
subset of the flows (\eg fewer than 1\% for coflows with many flows).  (2)
Since the CCT of a coflow depends on the completion of its last flow, some of
its earlier finishing flows could be delayed without affecting the CCT. \name
exploits this observation and schedules pilot flows on the least-busy ports to
increase the odds that it only affects earlier finishing flows of other
coflows.

Second, scheduling pilot flows first may elongate the CCT of the newly arriving
coflow itself whose other flows cannot start until the pilot flows finish. This
is again typically insignificant for two reasons: (1) A coflow (\eg from a
MapReduce job) typically consists of flows from all sending ports to all
receiving ports. Conceptually, pre-scheduling one out of multiple flows from
each sender may not delay the coflow progress at that port, because all flows
at that port have to be sent anyway.
(2) Coflow scheduling is of high relevance in a busy cluster (when there is a
backlog of coflows in the network), in which case the CCT of coflow is expected
to be much higher than if it were the only coflow in the network, and hence the
piloting overhead is further dwarfed by a coflow's actual CCT.

\vspace{-0.1in}
\subsection{Why is sampling effective in the presence of skew?}

The flow sizes within a coflow may vary (\emph{skew}). \addtonreview{In this paper we measure skew as $\frac{max\ flow\ length}{min\ flow\ length}$. Other papers like Varys\cite{varys:sigcomm14} have used metrics like coefficient of variation to measure the skew. We used the ratio $\frac{max\ flow\ length}{min\ flow\ length}$ because it allows us to analyze the learning error without assuming the specific distribution of flow-sizes.} 

Intuitively, if the skew across flow sizes is small, sampling even a small
number of pilot flows will be sufficient to yield an accurate estimate.
Interestingly, even if the skew across flow sizes is large, our experiment
indicates that sampling is still highly effective.
In the following, we give both the intuition and theoretical underpinning for why
sampling is effective.

{Consider, for example, two coflows and the simple setting where both coflows
share the same set of ports}. In order to improve the average CCT, we wish to
schedule the shorter coflow ahead of the longer coflow. If the total sizes of
the two coflows are very different, then even a moderate amount of estimation
error of the coflow sizes will not alter their ordering. On the other hand, if
the total sizes of the two coflows are close to each other, then indeed the
estimation errors will likely alter their ordering. However, in this case since
their sizes are not very different anyway, switching the order of these two
coflows will not significantly affect the average CCT. 

\paragraph{Analytic results.}
To illustrate the above effect, we show that the gap between the CCT based on
sampling and assuming perfect knowledge is small, even under general flow size
distributions.  Specifically, coflows $\coflowone$  and $\coflowtwo$  have $c
\nnone$ and $c \nntwo$ flows, respectively. Here, we assume that $\nnone$ and
$\nntwo$ are fixed constants. Thus, by taking $c$ to be larger, we will be able
to consider wider coflows.  
Assume that each flow of $\coflowone$ (correspondingly, $\coflowtwo$) has a
size that is distributed within a bounded interval [$a_1, b_1$] ([$a_2, b_2$])
with mean $\muone$ ($\mutwo$), \emph{i.i.d.} across flows.
Let $\Tc$ be the total completion time when the exact flow sizes are known in
advance. Let $\Tcprime$ be the average CCT by sampling $\mone$ and $\mtwo$
flows from $\coflowone$ and $\coflowtwo$, respectively. Without loss of
generality, we assume that $\nntwo \mutwo \ge \nnone \muone$.


Then, using Hoeffding's Inequality, we can show that (see
\S\ref{sec:appendix} for detailed proof)
{\small
\begin{equation}
\label{eqn:general} 
\begin{split}
\lim_{c \to \infty} \frac{\Tcprime-\Tc}{\Tc} \le
4 \exp\left[-\frac{2 (\nntwo \mutwo - \nnone \muone)^2}
	{\left( \frac{\nntwo (b_2-a_2)}{\sqrt{\mtwo}} 
		+ \frac{\nnone (b_1-a_1)}{\sqrt{\mone}} \right)^2}
	\right] \\
\frac{\nntwo \mutwo - \nnone \muone} {\nntwo \mutwo + 2 \nnone \muone }
\end{split}
\end{equation} 
}
{(Note that here also we have used the fact that, since both
coflows share the same set of ports and $c$ is large, 
the CCT is asymptotically proportional to the coflow
size.)}

Equation (\ref{eqn:general}) can be interpreted as follows. First, due to the
first exponential term, the relative gap between $\Tcprime$ and $\Tc$ decreases
as $b_1 - a_1$ and $b_2 - a_2$ decrease. In other words, as the skew of each
coflow decreases, sampling becomes more effective.  Second, when $b_1 - a_1$
and $b_2 - a_2$ are fixed, if $\nntwo \mutwo - \nnone \muone$ is large (i.e.,
the two coflow sizes are very different), the value of the exponential function
will be small. On the other hand, if $\nntwo \mutwo - \nnone \muone$ is close
to zero (i.e., the two coflow sizes are close to each other), the numerator on
the second term on the right hand side will be small. In both cases, the
relative gap between $\Tcprime$ and $\Tc$ will also be small, which is
consistent with the intuition explained earlier. The largest gap occurs when
$\nntwo \mutwo - \nnone \muone$ is on the same order as $\frac{\nntwo
(b_2-a_2)}{\sqrt{\mtwo}} + \frac{\nnone (b_1-a_1)}{\sqrt{\mone}}$. 
Finally, although these analytical results assume that both
coflows share the same set of ports, similar conclusions on
the impact of estimation errors due to sampling also apply under
more general settings.

{{The above analytical results suggest that, when $c$
		is large, the relative performance gap for CCT is
		a function of the number of pilot flows sampled for each
		coflow, but is independent of the total number of flows  
		in each coflow. In practice, large coflows will
		dominate the total CCT in the system. Thus, these
		results partly explain that, while in our experiments
		the number of pilot flows is never larger than $1\%$ of
		the total number of flows, the performance of our
		proposed approach is already very good.
		
                Finally, the above analytical
		results do not directly tell us how to choose the
		number of pilot flows, which likely depends on the
		probability distribution of the flow size. In practice, we
		do not know such distribution ahead of time. Further,
		while choosing a larger number of pilot flows reduces 
		the estimation errors, it also incurs higher overhead
		and delay. Therefore, our design (\S\ref{sec:design}) needs to have
		practical solutions that carefully address these
issues.}} 


\linx{
\textbf{Error-correction:}
Readers familiar with the online
	learning and multi-armed
	bandit (MAB) literature 
\cite{gittins2011multi,bubeck2012regret,lai1985asymptotically,auer2002finite}
will notice that our key idea above does not attempt to correct the 
errors in the sampling step. In such literature, it is common to use
an iterative procedure so that more samples are collected to correct any
initial errors. Such an iterative procedure must 
carefully balance
``exploration'' (i.e., an arm that looks poor at the beginning may
actually have high expected payoff in the long term)
and ``exploitation'' (i.e., arms that already demonstrate high
payoffs should be used more often). 
For instance, the Upper-confidence-bound (UCB) algorithm of 
\cite{auer2002finite} is a classical example that adjusts the sampled mean by a
confidence interval that decreases with the number of
samples, so that preference is given to arms that have only been
sampled a small number of times, which may thus have large errors. In this
way, UCB has been shown to balance exploitation and exploration. 
\emph{However, in this paper we have not used {this type of} ideas due
to a fundamental difference in our objective.} Specfically, our objective is 
to minimizing the average CCT, instead of maximizing the total payoff.
To see the difficulty of UCB-type of algorithms, consider the case when
we have two Coflows whose sizes are nearly identical. 
A straight-forward way of using {confidence} intervals is to compute a
lower-confidence-bound for the sampled mean, i.e., by subtracting the
size of the confidence interval from the sampled mean\footnote{\linx{It is
	more logical to use a lower-confidence-bound here because we aim to
	\emph{minimize} the completion time, but not to \emph{maximize} the
payoffs.}}.
In this way, preference is given to the Coflow with a smaller size
(according to the sampled mean) or with a smaller number of samples.
Then, as new samples of the Coflow are revealed, we recompute the
confidence interval and reschedule all remaining flows
accordingly. Under such a scheme, suppose that Coflow $\coflowone$ initially has a
smaller sampled mean than Coflow $\coflowtwo$,
but the sizes of their confidence intervals are initially the same. Thus,
$\coflowone$ will be scheduled first. As new samples of $\coflowone$ becomes
available, its confidence interval shrinks. It is then likely that
$\coflowtwo$ will have a smaller lower-confidence-bound, and thus will be
scheduled next. In this way, $\coflowone$ and $\coflowtwo$ are essentially scheduled
in a round-robin manner because they turns becoming the one with
the smaller lower-confidence-bound. As a result, they will both complete
at about the same time. In contrast, a better strategy in this situation
is to let one Coflow runs until completion, before the other Coflow
starts. Due to this reason, in this paper we do not use confidence
intervals nor an iterative procedure to correct sampling errors. 
Nonetheless, thanks to the reasons explained in main article 
by 
$eq.$(1)
, we have observed that
our proposed approach already produces superior results.}

To verify the above intuition, we have conducted the following experiments with
adding error correction in size estimation module (\S IV-B)
the default \name (\S IV)
. For this we need (1) a method to calculate
the confidence interval on the estimated size; and (2) a method for error
correction.  In our experiments to evaluate \name with error correction we use
the following methods. (1) We calculate confidence interval of the estimated
size by bootstraping~\cite{bootstrapMethods}. Specifically, we resample 100
times from the same set of sampled flows (with repitition allowed) and then
using the 100 averages calculate the lower bound of the confidence interval as
$average - 3*standard\_deviation$.  (2) The error correction process triggers
each time a specific set of flows complete in the following manner. Each set
has the same number of flows as the number of pilot
flows (\S IV-A)
, say $p$ flows, for the Coflow. These sets are
ordered according to the time that the corresponding flows start, \ie first $p$
flows that start after size estimation is over are grouped as the first set of
error-correcting flows and the next $p$ flows that start are grouped as second
set, and so on. We perform the first round of error-correction (\ie
recalculating the lower confidence bound using the method (1) (described above))
only after all flows from the first set are completed, and the second round
after all flows from the second set are completed. In this way we avoid any
sort of bias of the samples due to completion time. Using the above two
features we evaluated three variants of the default \name (\S IV)
against Aalo based on the FB-Coflow trace.
(1) \name-lower-confidence-bound: this variant uses the lower bound of the
confidence interval of estimated size of a Coflow as its estimated size (Note
that in contrast the default \name uses the unbiased mean size); (2)
\name-with-one-error-correction: on top of (1), we apply one round of error
correction, after first set of $p$ error-correction flows are completed, so
that more flow sizes are used to update the estimate; (3)
\name-with-multiple-error-correction: on top of (1), error correction is
applied multiple rounds till the Coflow finishes.  Our evaluation shows that
variant (1) improves the average CCT over Aalo by 1.33$\times$ and median (P90)
CCT speedup by 1.78$\times$ (10.75$\times$), variant (2) improves average CCT by
1.27$\times$ and median (P90) CCT speedup by 1.59$\times$ (9.78$\times$), and
variant (3) degrades the average CCT by 0.95$\times$ and median (P90) CCT
speedup is 1.06$\times$ (8.25$\times$). However, the default \name (without
error correction) improves average CCT by 1.51$\times$ and median (P90) CCT
speedup by 1.78$\times$ (9.58$\times$) (\S VIII-D)
. In summary, adding
error-correction only seems to degrade the performance of \name in all 3
variants

\linx{We note that this result does not preclude the possibilities that other
iterative sampling algorithms may outperform Philae. Nonetheless, it illustrates
that straight-forward extensions of UCB-type of ideas do not work well.
How to find iterative sampling algorithms outperforming Philae remains
an interesting direction for future work.}
\subsection{Scalability analysis}
\label{sec:scalability}

Compared to learning coflow sizes using priority
queues (PQ-based)~\cite{aalo:sigcomm15,jajooSaath}, learning coflow sizes by sampling \name
not only reduces the learning overhead as discussed in \S\ref{subsec:why}
and shown in \S\ref{sec:sim:probeOverhead}, but also significantly
reduces the amount of interactions between the coordinator and local
agents and thus makes the coordinator highly scalable, as summarized
in Table~\ref{table:scalability:updatesummary}.

{First, {\em PQ-based learning requires much more frequent update from local agents}.}
PQ-based learning estimates coflow sizes by incrementally moving coflows across
priority queues according to the data sent by them so far. As such,
the scheduler needs frequent updates (every $\delta$ ms) of data sent per coflow from the local
agents.
In contrast, \name directly estimates a coflow's size upon the
completion of all its pilot flows. The only updates \name needs from
the local agents are about the flow completion which is needed for
{updating contentions and removing flows from active consideration.}.

{Second, {\em PQ-based learning results in much more frequent rate allocation}.}
In sampling-based approach, since coflow sizes are estimated only
once, coflows are re-ordered only upon {coflow completion or
arrival events} or in the case of contention based policies only when
contention changes, which is {triggered by completion of all the flows of a
  coflow at a port}.  In contrast, in PQ-based learning,
at every $\delta$ interval, coflow data sent are updated and coflow priority may get updated,
which will trigger new rate assignment.

Our scalability experiments in \S\ref{sec:eval:scale} confirms that \name
achieves much higher scalability than Aalo.

\begin{table}[tp]
\caption{Comparison of frequency of interactions between the coordinator and local agents.}
\label{table:scalability:updatesummary}
\centering
{\small
\begin{tabular}{|c|c|c|c|c|c|}
\hline
	& Update& Update of & Rate\\
	& of data sent & flow completion & calculation \\
\hline
	\name & No & Yes & Event triggered\\
\hline
	Aalo & Periodic ($\delta$)& Yes & Periodic ($\delta$)\\
\hline
\end{tabular}
}
\vspace{-0.2in}
\end{table}

\section{Implementation}
\label{sec:impl}

We implemented both \name and Aalo scheduling policies in the same framework
consisting of the global coordinator and local agents
(Fig.~\ref{fig:design:arch}), in 5.2 KLoC in C++.


\textbf{Coordinator:} The coordinator schedules the coflows based on the
operations received from the registering framework. The key implementation challenge for
the coordinator is that it needs to be fast in computing and updating the
schedules. The \name coordinator is optimized for speed using a variety of
techniques including pipelining, process affinity, and concurrency whenever
possible.

\textbf{Local agents:} The local agents update the global coordinator only 
upon completion of a flow,
along with its length if it is a pilot flow. Local agents schedule the
coflows based on the last schedule received from the coordinator. They
comply to the last schedule until a new schedule is received. To
intercept the packets from the flows, local agents require the compute
framework to replace {\small \texttt{datasend(), datarecv()}} APIs with the
corresponding \name APIs, which incurs very small overhead.

\textbf{Coflow operations:} The global coordinator runs independently from, and
is not coupled to, any compute framework, which makes it general enough to be
used with any framework.  It provides RESTful APIs to the frameworks for coflow
operations: (a) {\small \texttt{register()}} for registering a new coflow when
it enters, (b) {\small \texttt{deregister()}} for removing a coflow when it
exits, and (c) {\small \texttt{update()}} for updating coflow status whenever
there is a change in the coflow structure, \eg during task migration
and restarts after node failures.





\vspace{-0.1in}

\vspace{-0.2in}
\section{Testbed Evaluation}
\label{sec:eval}
Next, we deployed \name in a 150-machine Azure cluster
and a 900-machine cluster to evaluate its performance and scalability.

\textbf{Testbed setup:}
We rerun the FB trace on a Spark-like framework on a
150-node cluster in Microsoft Azure~\cite{azure:web}.  The coordinator runs on
a Standard DS15 v2 server with 20-core 2.4 GHz Intel Xeon E5-2673 v3 (Haswell)
processor and 140GB memory. The local agents run on D2v2 with the same processor
as the coordinator
with 2-core and 7GB memory. The machines on which local agents run have 1 Gbps
network bandwidth.  Similarly as in simulations, our testbed evaluation keeps the
same flow lengths and flow ports in trace replay. All the experiments use
default parameters $K, E, S$ and the default pilot flow selection policy.

\vspace{-0.1in}
\subsection{CCT Improvement}

In this experiment, we measure CCT improvements of \name compared to Aalo.
Fig.~\ref{fig:eval:jct-cct-speedUp} shows the CDF of the CCT speedup of
individual coflows under \name compared to under Aalo. The average CCT improvement is
1.50$\times$ which is similar to the results in the simulation experiments.
We also observe 1.63$\times$ P50 speedup and 8.00$\times$ P90 speedup.

We also evaluated \name using the Wide-coflow-only trace.
Table~\ref{table:eval:cct} shows that \name achieves 1.52$\times$ improvement
in average CCT over Aalo, similar to that using the full FB trace.
{This is because the improvement in average CCT is dominated by large
  coflows, \name is speeding up large coflows,
  and the Wide-coflow-only trace consists of mostly large coflows.}

\begin{table}[tp]
	\caption{[Testbed] CCT improvement in \name as compared to Aalo.}
\vspace{-0.1in}
\label{table:eval:cct}
\centering
{\small
\begin{tabular}{|c|c|c|c|}
\hline
  & P50 & P90 & Avg. CCT \\
\hline
FB Trace  & 1.63$\times$  &8.00$\times$ &1.50$\times$ \\
\hline
Wide-coflow-only  & 1.05$\times$ & 2.14$\times$ & 1.49$\times$ \\
\hline
\end{tabular}
}
\vspace{-0.15in}
\end{table}

\vspace{-0.1in}
\subsection{Job Completion Time}
\label{sec:eval:jct}

Next, we evaluate how the improvement in CCT affects the job completion time
(JCT).  In data clusters, different jobs spend different fractions of their
total job time in data shuffle.
In this experiment, we used 526 jobs, each corresponding to one
coflow in the FB trace. The fraction of time that the jobs 
spent in the shuffle phase follows the same distribution used in
Aalo~\cite{aalo:sigcomm15}, \ie 61\% jobs spent less than 25\% of their
total time in shuffle, 13\% jobs spent 25-49\%, another 14\% jobs spent 50-74\%, and
the remaining spent over 75\% of their total time in shuffle.
Fig.~\ref{fig:eval:jct-cct-speedUp} shows the
CDF of individual speedups in JCT.  Across all jobs, \name reduces the job
completion time by 1.16$\times$ in the median case and 7.87$\times$ in the
90$^{th}$ percentile.
This shows that improved CCT translates into better job completion time. As
expected, the improvement in job completion time is smaller than the
improvement in CCT because job completion time depends on the time spent in
both compute and shuffle (communication) stages, and \name improves only the
communication stage.

\vspace{-0.1in}
\subsection{Scalability}
\label{sec:eval:scale}

\setlength\tabcolsep{2.25pt}
\begin{table}[tp]
\caption{[Testbed] Average (standard deviation) coordinator CPU time (ms)
per scheduling interval in 900-port runs.
\name did not have to calculate and send new rates in 66\% of intervals,
which contributes to its low average.
}
\label{table:eval:scale:profile}
\centering
{\small
\vspace{-0.1in}
\begin{tabular}{|c|c|c|c|c|c|}
\hline
	& Rate Calc. & New Rate Send & Update Recv. & Total\\
\hline
	\name & 2.99 (5.35)& 4.90 (11.25)& 6.89 (17.78)& 14.80 (28.84)\\
\hline
	Aalo & 4.28 (4.14)& 17.65 (20.90)& 10.97 (19.98)& 32.90 (34.09)\\
\hline
\end{tabular}
}
\vspace{-0.2in}
\end{table}
\begin{table}[tp]
\caption{[Testbed] Percentage of scheduling intervals where
synchronization and rate calculation took
	more than $\delta$ for 150-port and $\delta'(= 6\times\delta)$ for 900-port runs.}
\label{table:eval:scale:miss}
\centering
{\small
\vspace{-0.1in}
\begin{tabular}{|c|c|c|c|c|c|}
\hline
	& 150 ports & 900 ports\\
\hline
	\name & 1\%& 10\%\\
\hline
	Aalo & 16\%& 37\%\\
\hline
\end{tabular}
}
\vspace{-0.1in}
\end{table}

Finally, we evaluate the scalability of \name by comparing its
performance with Aalo on a 900-node cluster. To drive the evaluation,
we derive a 900-port trace by replicating the FB trace 6 times across
ports, \ie we replicated each job 6 times, keeping the arrival time
for each copy the same but assigning sending and receiving ports in
increments of 150 (the cluster size for the original trace). We also
increased the scheduling interval $\delta$ by 6 times to 
$\delta'$ = 6$\times$$\delta$.

\name achieved 2.72$\times$ (9.78$\times$) speedup in average (P90) CCT over Aalo. 
The higher speedup compared to the 150-node runs (1.50$\times$) comes
from higher scalability of \name.  In 900-node runs, Aalo was not able
to finish receiving updates, calculating new rates and updating local
agents of new rates within $\delta'$ in 37\% of the intervals,
whereas \name only missed the deadline in 10\% of the intervals. For
150-node runs these values are 16\% for Aalo and 1\% for \name. The
21\% increase in missed scheduling intervals in 900-node runs in Aalo
resulted in local agents executing more frequently with outdated
rates. As a result, \name achieved even higher speedup in 900-node runs.

As discussed in \S\ref{sec:scalability}, Aalo's poorer coordinator
scalability comes from more frequent updates from local agents and
more frequent rate allocation, which result in longer coordinator CPU
time in each scheduling interval.
Table~\ref{table:eval:scale:profile} shows the average coordinator CPU
usage per interval and its breakdown.
We see that (1) on average \name spends
much less time than Aalo in receiving updates from local agents,
because \name does not need updates from local agents at every interval --
on average in every
scheduling interval \name receives updates from 49 local agents whereas Aalo
receives from 429 local agents,
and (2) on average \name 
spends much less time calculating new rates and send new rates.  This is
because rate calculation in \name is triggered by events and \name did
not have to flush rates in 66\% of the intervals.

\iftoggle{tons}{
{
\vspace{-0.15in}
\subsection{Robustness to network error}
\label{sec:eval:networkError}

As discussed in \S\ref{sec:impl}, unlike Aalo, \name's coordinator does not
need constant updates from local agents to sort coflows in priority queues.
This simplifies \name's design and makes it robust to network error. To
evaluate the benefit of this property of \name, we evaluated Aalo and \name 5
times with the same configuration {using the FB trace}.
{Table~\ref{table:eval:networkRobust}
shows the mean-normalized standard deviation in the 10$^{th}$, 50$^{th}$,
90$^{th}$ percentile and the average CCT across the 5 runs.}
The lower values for \name indicates that it is more robust to network
dynamics than Aalo.

\begin{table}[tp]
\caption{[Testbed] Mean normalised standard deviation in CCT among
 \name and Aalo.}
\label{table:eval:networkRobust}
\centering
\vspace{-0.1in}
{\small
\begin{tabular}{|c|c|c|c|c|}
\hline
 & P10 & P50 & P90 & Avg. CCT \\
\hline
\name &6.1\%  & 2.3\%  &0.1\%  &0.1\%  \\
\hline
Aalo & 7.1\% & 4.4\% & 2.7\% & 1.6\% \\
\hline
\end{tabular}
}
\vspace{-0.1in}
\end{table}
%
\vspace{-0.15in}
\subsection{Resource Utilization}
\label{sec:eval:resource}

Finally, we evaluate, for both \name and Aalo, the resource utilization at the
coordinator and the local agents (Table~\ref{table:eval:resource}) in terms of
CPU and memory usage.  We measure the overheads in two cases: (1) Overall:
average during the entire execution of the trace, (2) Busy: the 90-th
percentile utilization indicating the performance during busy periods due to a
large number of coflows arriving. As shown in Table~\ref{table:eval:resource},
\name agents have similar utilization as Aalo at the local nodes, where the CPU
and memory utilization are minimal even during busy times.  The global
coordinator of \name consumes much lower server resources than Aalo -- the CPU
utilization is 3.4$\times$ lower than Aalo on average, and 2.6$\times$ than
Aalo during busy periods.  This is due to \name's event triggered communication
and sampling-based learning, which significantly lowers its communication
frequency with local agents when compared to Aalo.  The lower resource
utilization of the global coordinator enables \name to scale to a lager cluster
size than Aalo.

\setlength\tabcolsep{2.25pt}
\begin{table}[tp]
	\caption{\hspace{-0.1in}[Testbed] Resource usage in \name and Aalo for 150 ports experiment.}
\vspace{-0.1in}
\label{table:eval:resource}
\centering
{\small
\begin{tabular}{|c|c|c|c|c|c|}
\hline
& & \multicolumn{2}{|c|}{\name} & \multicolumn{2}{|c|}{Aalo} \\
\hline
 & & Overall & Busy & Overall & Busy \\
\hline
\hline
Coordi- & CPU (\%) & 5.0 & 10.4 & 17.0 & 27.2  \\
nator & Memory (MB) & 212 & 218 & 318 & 427 \\
\hline
Local & CPU (\%) & 4.3 & 4.6 & 4.5  & 4.6    \\
node & Memory (MB) &1.65  & 1.70  &1.64  &1.70  \\
\hline
\end{tabular}
}
\vspace{-0.2in}
\end{table}
}
}{}
\vspace{-0.1in}
\section{Related Work}
\label{sec:related}
\textbf{Coflow scheduling:}
In this paper, we have shown \name outperforms prior-art non-clairvoyant coflow
scheduler Aalo from more efficient learning of coflow sizes online.
In~\cite{aalo:sigcomm15}, Aalo was shown to outperform previous non-clairvoyant
coflow schedulers Baraat~\cite{baraat:sigcomm14} by using global coordination,
and Orchestra~\cite{orchestra:sigcomm11} by avoiding head-of-line blocking.

Clairvoyant coflow schedulers such as Varys~\cite{varys:sigcomm14} and
{Sincronia~\cite{sincronia:sigcomm18}} assume prior knowledge of
coflows upon arrival.
Varys runs a shortest-effective-bottleneck-first heuristic for
inter-coflow scheduling and performs per-flow rate allocation at the
coordinator.  Sincronia improves the scalability of the centralized
coordinator of Varys by only calculating the coflow ordering at the
coordinator (by solving an LP) and offloading flow rate allocation to
individual local agents. Sincronia is orthogonal
to \name; once coflow sizes are learned through sampling, ideas from
Sincronia can be adopted in \name to order coflows and offload rate
allocation to local ports.
\commenttons{Prioritized work conservation in Sincronia helps in mitigating the effect
of starvation, which arises as a result of ordering coflows in
the optimal order. However, it still does not guarantee complete freedom
from starvation.}
CODA~\cite{zhang:sigcomm15} tackles an orthogonal problem
of identifying flows of individual coflows online.

However, recent studies~\cite{plausibleFlowSize, aalo:sigcomm15} have shown
various reasons why it is not very plausible to learn flow sizes from
applications beforehand. For example, many applications stream data as soon as
data are generated and thus the application does not know the flow sizes until
flow completion, and learning flow sizes from applications requires changing
either the network stack or the applications.


\textbf{Flow scheduling:} There exist a rich body of prior work on flow
scheduling. Efforts to minimize flow completion time (FCT), both with prior
information 
(pFabric~\cite{pfabric:sigcomm13})
and without prior information (\eg Fastpass~\cite{fastpass:sigcomm14})
, fall short in minimizing CCTs
which depend on the completion of the last flow~\cite{varys:sigcomm14}.
Similarly, Hedera~\cite{hedera:nsdi10} and MicroTE~\cite{microte:conext11}
schedule the flows with the goal of reducing the overall FCT, which again is
different from reducing the overall CCT of coflows.

\textbf{Speculative scheduling} Recent works~\cite{creditscheduling:sigcomm17,
trumpet:sigcomm16} use the idea of online requirement estimation for
scheduling in datacenter. In~\cite{corral:sigcomm15}, recurring big data
analytics jobs are scheduled using their history.

\textbf{Job scheduling:} There have been much work on scheduling in analytic
systems and storage at scale by improving speculative tasks~\cite{late:osdi08,
mantri:osdi10, dolly:nsdi13, slearnTechReport, jajooSLearn,
jajoo2020exploiting}, improving locality~\cite{delay:eurosys10,
scarlett:eurosys11}, and end-point
flexibility~\cite{sinbad:sigcomm13, pisces:osdi12}. The Coflow
abstraction is complimentary to these work, and can benefit from them. 
Combining Coflow with these approaches remains as a future work.

\textbf{Job scheduling:} There have been much work on scheduling in analytic
systems and storage at scale by improving speculative tasks~\cite{late:osdi08,
mantri:OSDI2010, dolly:nsdi13}, improving locality~\cite{delay:eurosys10,
scarlett:eurosys11}, and end-point
flexibility~\cite{sinbad:sigcomm13, pisces:osdi12}. The coflow
abstraction is complimentary to these work, and can benefit from them. 
Combining coflow 
them 
remains a future work.

\textbf{Scheduling in parallel processors:} Coflow scheduling by exploiting the
spatial dimension bears similarity to scheduling processes on parallel
processors and multi-cores, where many variations of
FIFO~\cite{cmpproc:soda98}, FIFO with backfilling~\cite{cmpproc:ipps95} and
gang scheduling~\cite{cmpproc:pp05}
have been proposed.


\vspace{-0.1in}
\section{Conclusion}
\label{sec:conc}

State-of-the-art online coflow schedulers approximate the classic SJF by
implicitly learning coflow sizes and pay a high penalty for large coflows. We
propose the novel idea of sampling in the spatial dimension of coflows to
explicitly and efficiently learn coflow sizes online to enable efficient online
SJF scheduling. Our extensive simulation and testbed experiments show the new
design offers significant performance improvement over prior art.
Further, the sampling-in-spatial-dimension technique can be
generalized to other distributed scheduling problems such as cluster
job scheduling.
We have made our simulator publicly
available at {\small {https://github.com/coflowPhilae/simulator}}~\cite{philaeSim}.




\bibliographystyle{plain}
\vspace{-0.1in}
\bibliography{coflow}
\pagenumbering{gobble}

\end{document}